\documentclass[a4paper,twocolumn,12pt]{article}
\usepackage[spanish]{babel}
\usepackage{ae}
\usepackage[latin1]{inputenc} 
\usepackage[T1]{fontenc} 
\usepackage{layout}
\usepackage{array}
\usepackage{graphicx}
\usepackage{amsmath}
\usepackage{multicol}
\usepackage{rotating}  
\hyphenrules{nohyphenation} 

\usepackage[]{cite}   
\usepackage{anysize}  
\usepackage{endnotes} 

\usepackage{tikz,fp,ifthen,fullpage}
\usetikzlibrary{arrows,snakes,backgrounds}
\usepackage{pgfplots}
\usetikzlibrary{shapes,trees}
\usetikzlibrary{calc,through,backgrounds,decorations}
\usetikzlibrary{decorations.shapes,decorations.text,decorations.pathmorphing,backgrounds,fit,calc,through,decorations.fractals}
\usetikzlibrary{fadings,intersections}
\usetikzlibrary{patterns}
\usetikzlibrary{mindmap}
\marginsize{1.5cm}{1.0cm}{1.5cm}{2cm}

\begin{document}

\renewcommand{\tablename}{Tabla}
\renewcommand{\abstractname}{}
\renewcommand{\thefootnote}{\arabic{footnote}}

\title{
        {\LARGE Aplicación del  FCI a estudiantes de ingeniería de Bogotá: una interpretación de los resultados mediante un modelo aleatorio de dos niveles}\\
        \footnotesize   \textit{(Application of FCI at engineering students in Bogota: an interpretation of the  answers through a random model of two levels)}
}
\author
{ 
Paco  Talero $^{1}$,Orlando Organista$^{1}$ y Luis Barbosa$^{1}$\\
$^{1}$ {\small Grupo F\'{\i}sica y Matem\'{a}tica, Dpt de Ciencias
Naturales, Universidad Central, Bogotá Colombia}
}
\date{}
\twocolumn
[
\begin{@twocolumnfalse}
\maketitle 
\begin{abstract}
Se aplicó el FCI a $646$ estudiantes de ingeniería de Bogotá justo al comenzar su primer curso de física, se encontró que la frecuencia relativa del número de preguntas correctas obedece a un modelo aleatorio de dos  niveles y no a modelos mentales correctos sobre el mundo físico.\\
\textbf{Palavras-chave:} Bogotá, FCI, modelo aleatorio.\\ 

We applied the FCI to $646$ engineering students from Bogota when they began your first year physics, we found that the relative frequency of the number of correct answers has a random pattern of two levels, also we found that they don´t have clear  mental models  about physical world.\\

\end{abstract}
\end{@twocolumnfalse}
]

\section{Introducción}
Desde su aparición en $1992$ el FCI (Force Concept Inventory)  ha sido una herramienta fundamental en diversas investigaciones en enseñanza de la física \cite{Mora,Compadre,LieB_1,LieB_2,Julio,Hestenes1}. El FCI es un cuestionario compuesto por $30$ preguntas de selección múltiple con una única respuesta correcta y cuatro opciones que corresponden al pensamiento común,  indaga por el pensamiento Newtoniano  a nivel conceptual y está dividido en seis  bloques: cinemática, primera ley, segunda ley, tercera ley, principio de superposición y los tipos de fuerza. Este instrumento ha sido adaptado y aplicado en varios países con diversos objetivos de investigación\cite{Mora,LieB_2}. Sin embargo, no se conocían estudios sobre el pensamiento Newtoniano indagados mediante el FCI en estudiantes de ingeniería de Bogotá Colombia.\\

En la sección ($2$) se detalla la aplicación del FCI en estudiantes de ingeniería de Bogotá y se muestra la frecuencia relativa del número de preguntas correctas, en la sección ($3$) se plantea el modelo aleatorio de dos respuestas y se confronta con los resultados obtenidos en el trabajo de campo, finalmente  en la sección ($4$) se muestran las conclusiones. 

\section{Trabajo de campo}
Con el objetivo de sondear el pensamiento Newtoniano de manera conceptual en estudiantes de ingeniería de Bogotá se aplicó  durante el primer semestre del año $2012$  el FCI \footnote{Traducción y adaptación de los profesores Genaro Zabala y Julio Venegas} a $786$ estudiantes de $8$ universidades que imparten algún programa de ingeniería. La aplicación del cuestionario se realizó con autorización formal de las directivas y profesores, se aplicó  tanto al  inicio como al final del primer curso de física impartido en cada universidad, el  cuestionario fue aplicado algunas veces por el profesor del curso o su monitor  y otras  directamente por los investigadores.\\    

Los resultados de la investigación arrojaron una fuerte heterogeneidad  marcada por la presencia de dos grupos de estudiantes con desempeños marcadamente diferentes. El primero constó de $140$ estudiantes en los cuales se observó alta coherencia en sus respuestas permitiendo tomar al FCI como un instrumento confiable  para indagar por las preconcepciones y la evolución del aprendizaje, el segundo se conformó de $646$ estudiantes que se caracterizó por exhibir respuestas con bajo puntaje y baja concentración que hacen del FCI un instrumento poco confiable desde el punto de vista de la teoría clase del test (TCT). Este artículo se enfoca en el estudio de algunas características de este último grupo.\\
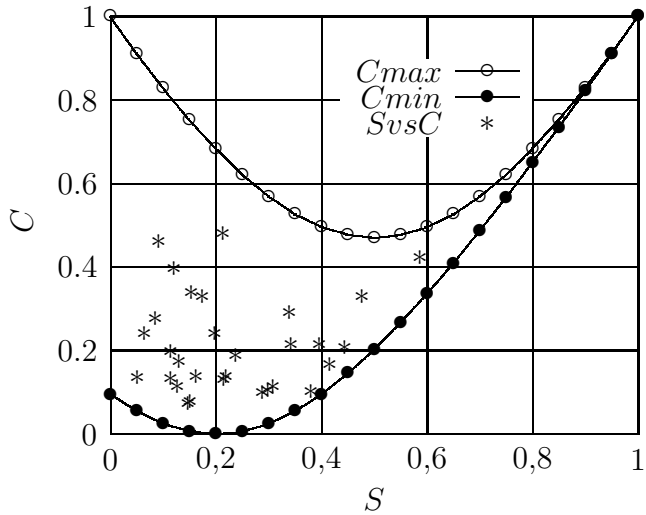
\begin {figure}
\begin{center}
\setlength{\unitlength}{0.240900pt}
\ifx\plotpoint\undefined\newsavebox{\plotpoint}\fi
\sbox{\plotpoint}{\rule[-0.200pt]{0.400pt}{0.400pt}}%
\begin{picture}(1062,826)(0,0)
\sbox{\plotpoint}{\rule[-0.200pt]{0.400pt}{0.400pt}}%
\put(191.0,131.0){\rule[-0.200pt]{197.538pt}{0.400pt}}
\put(191.0,131.0){\rule[-0.200pt]{4.818pt}{0.400pt}}
\put(171,131){\makebox(0,0)[r]{ 0}}
\put(991.0,131.0){\rule[-0.200pt]{4.818pt}{0.400pt}}
\put(191.0,262.0){\rule[-0.200pt]{197.538pt}{0.400pt}}
\put(191.0,262.0){\rule[-0.200pt]{4.818pt}{0.400pt}}
\put(171,262){\makebox(0,0)[r]{ 0.2}}
\put(991.0,262.0){\rule[-0.200pt]{4.818pt}{0.400pt}}
\put(191.0,393.0){\rule[-0.200pt]{197.538pt}{0.400pt}}
\put(191.0,393.0){\rule[-0.200pt]{4.818pt}{0.400pt}}
\put(171,393){\makebox(0,0)[r]{ 0.4}}
\put(991.0,393.0){\rule[-0.200pt]{4.818pt}{0.400pt}}
\put(191.0,523.0){\rule[-0.200pt]{197.538pt}{0.400pt}}
\put(191.0,523.0){\rule[-0.200pt]{4.818pt}{0.400pt}}
\put(171,523){\makebox(0,0)[r]{ 0.6}}
\put(991.0,523.0){\rule[-0.200pt]{4.818pt}{0.400pt}}
\put(191.0,654.0){\rule[-0.200pt]{85.760pt}{0.400pt}}
\put(847.0,654.0){\rule[-0.200pt]{39.508pt}{0.400pt}}
\put(191.0,654.0){\rule[-0.200pt]{4.818pt}{0.400pt}}
\put(171,654){\makebox(0,0)[r]{ 0.8}}
\put(991.0,654.0){\rule[-0.200pt]{4.818pt}{0.400pt}}
\put(191.0,785.0){\rule[-0.200pt]{197.538pt}{0.400pt}}
\put(191.0,785.0){\rule[-0.200pt]{4.818pt}{0.400pt}}
\put(171,785){\makebox(0,0)[r]{ 1}}
\put(991.0,785.0){\rule[-0.200pt]{4.818pt}{0.400pt}}
\put(191.0,131.0){\rule[-0.200pt]{0.400pt}{157.549pt}}
\put(191.0,131.0){\rule[-0.200pt]{0.400pt}{4.818pt}}
\put(191,90){\makebox(0,0){$0$}}
\put(191.0,765.0){\rule[-0.200pt]{0.400pt}{4.818pt}}
\put(355.0,131.0){\rule[-0.200pt]{0.400pt}{157.549pt}}
\put(355.0,131.0){\rule[-0.200pt]{0.400pt}{4.818pt}}
\put(355,90){\makebox(0,0){$0.2$}}
\put(355.0,765.0){\rule[-0.200pt]{0.400pt}{4.818pt}}
\put(519.0,131.0){\rule[-0.200pt]{0.400pt}{157.549pt}}
\put(519.0,131.0){\rule[-0.200pt]{0.400pt}{4.818pt}}
\put(519,90){\makebox(0,0){$0.4$}}
\put(519.0,765.0){\rule[-0.200pt]{0.400pt}{4.818pt}}
\put(683.0,131.0){\rule[-0.200pt]{0.400pt}{112.259pt}}
\put(683.0,720.0){\rule[-0.200pt]{0.400pt}{15.658pt}}
\put(683.0,131.0){\rule[-0.200pt]{0.400pt}{4.818pt}}
\put(683,90){\makebox(0,0){$0.6$}}
\put(683.0,765.0){\rule[-0.200pt]{0.400pt}{4.818pt}}
\put(847.0,131.0){\rule[-0.200pt]{0.400pt}{157.549pt}}
\put(847.0,131.0){\rule[-0.200pt]{0.400pt}{4.818pt}}
\put(847,90){\makebox(0,0){$0.8$}}
\put(847.0,765.0){\rule[-0.200pt]{0.400pt}{4.818pt}}
\put(1011.0,131.0){\rule[-0.200pt]{0.400pt}{157.549pt}}
\put(1011.0,131.0){\rule[-0.200pt]{0.400pt}{4.818pt}}
\put(1011,90){\makebox(0,0){$1$}}
\put(1011.0,765.0){\rule[-0.200pt]{0.400pt}{4.818pt}}
\put(191.0,131.0){\rule[-0.200pt]{0.400pt}{157.549pt}}
\put(191.0,131.0){\rule[-0.200pt]{197.538pt}{0.400pt}}
\put(1011.0,131.0){\rule[-0.200pt]{0.400pt}{157.549pt}}
\put(191.0,785.0){\rule[-0.200pt]{197.538pt}{0.400pt}}
\put(50,458){\makebox(0,0){$\rotatebox{90} { \parbox{4cm}{\begin{align*}  C  \end{align*}}}$}}
\put(601,29){\makebox(0,0){$S$}}
\put(707,700){\makebox(0,0)[r]{$Cmax$}}
\put(727.0,700.0){\rule[-0.200pt]{24.090pt}{0.400pt}}
\put(191,785){\usebox{\plotpoint}}
\multiput(191.58,782.24)(0.498,-0.708){79}{\rule{0.120pt}{0.666pt}}
\multiput(190.17,783.62)(41.000,-56.618){2}{\rule{0.400pt}{0.333pt}}
\multiput(232.58,724.40)(0.498,-0.659){79}{\rule{0.120pt}{0.627pt}}
\multiput(231.17,725.70)(41.000,-52.699){2}{\rule{0.400pt}{0.313pt}}
\multiput(273.58,670.56)(0.498,-0.610){79}{\rule{0.120pt}{0.588pt}}
\multiput(272.17,671.78)(41.000,-48.780){2}{\rule{0.400pt}{0.294pt}}
\multiput(314.58,620.76)(0.498,-0.548){79}{\rule{0.120pt}{0.539pt}}
\multiput(313.17,621.88)(41.000,-43.881){2}{\rule{0.400pt}{0.270pt}}
\multiput(355.00,576.92)(0.499,-0.498){79}{\rule{0.500pt}{0.120pt}}
\multiput(355.00,577.17)(39.962,-41.000){2}{\rule{0.250pt}{0.400pt}}
\multiput(396.00,535.92)(0.603,-0.498){65}{\rule{0.582pt}{0.120pt}}
\multiput(396.00,536.17)(39.791,-34.000){2}{\rule{0.291pt}{0.400pt}}
\multiput(437.00,501.92)(0.734,-0.497){53}{\rule{0.686pt}{0.120pt}}
\multiput(437.00,502.17)(39.577,-28.000){2}{\rule{0.343pt}{0.400pt}}
\multiput(478.00,473.92)(1.033,-0.496){37}{\rule{0.920pt}{0.119pt}}
\multiput(478.00,474.17)(39.090,-20.000){2}{\rule{0.460pt}{0.400pt}}
\multiput(519.00,453.92)(1.746,-0.492){21}{\rule{1.467pt}{0.119pt}}
\multiput(519.00,454.17)(37.956,-12.000){2}{\rule{0.733pt}{0.400pt}}
\multiput(560.00,441.93)(4.495,-0.477){7}{\rule{3.380pt}{0.115pt}}
\multiput(560.00,442.17)(33.985,-5.000){2}{\rule{1.690pt}{0.400pt}}
\multiput(601.00,438.59)(4.495,0.477){7}{\rule{3.380pt}{0.115pt}}
\multiput(601.00,437.17)(33.985,5.000){2}{\rule{1.690pt}{0.400pt}}
\multiput(642.00,443.58)(1.746,0.492){21}{\rule{1.467pt}{0.119pt}}
\multiput(642.00,442.17)(37.956,12.000){2}{\rule{0.733pt}{0.400pt}}
\multiput(683.00,455.58)(1.033,0.496){37}{\rule{0.920pt}{0.119pt}}
\multiput(683.00,454.17)(39.090,20.000){2}{\rule{0.460pt}{0.400pt}}
\multiput(724.00,475.58)(0.734,0.497){53}{\rule{0.686pt}{0.120pt}}
\multiput(724.00,474.17)(39.577,28.000){2}{\rule{0.343pt}{0.400pt}}
\multiput(765.00,503.58)(0.603,0.498){65}{\rule{0.582pt}{0.120pt}}
\multiput(765.00,502.17)(39.791,34.000){2}{\rule{0.291pt}{0.400pt}}
\multiput(806.00,537.58)(0.499,0.498){79}{\rule{0.500pt}{0.120pt}}
\multiput(806.00,536.17)(39.962,41.000){2}{\rule{0.250pt}{0.400pt}}
\multiput(847.58,578.00)(0.498,0.548){79}{\rule{0.120pt}{0.539pt}}
\multiput(846.17,578.00)(41.000,43.881){2}{\rule{0.400pt}{0.270pt}}
\multiput(888.58,623.00)(0.498,0.610){79}{\rule{0.120pt}{0.588pt}}
\multiput(887.17,623.00)(41.000,48.780){2}{\rule{0.400pt}{0.294pt}}
\multiput(929.58,673.00)(0.498,0.659){79}{\rule{0.120pt}{0.627pt}}
\multiput(928.17,673.00)(41.000,52.699){2}{\rule{0.400pt}{0.313pt}}
\multiput(970.58,727.00)(0.498,0.708){79}{\rule{0.120pt}{0.666pt}}
\multiput(969.17,727.00)(41.000,56.618){2}{\rule{0.400pt}{0.333pt}}
\put(191,785){\makebox(0,0){$\circ$}}
\put(232,727){\makebox(0,0){$\circ$}}
\put(273,673){\makebox(0,0){$\circ$}}
\put(314,623){\makebox(0,0){$\circ$}}
\put(355,578){\makebox(0,0){$\circ$}}
\put(396,537){\makebox(0,0){$\circ$}}
\put(437,503){\makebox(0,0){$\circ$}}
\put(478,475){\makebox(0,0){$\circ$}}
\put(519,455){\makebox(0,0){$\circ$}}
\put(560,443){\makebox(0,0){$\circ$}}
\put(601,438){\makebox(0,0){$\circ$}}
\put(642,443){\makebox(0,0){$\circ$}}
\put(683,455){\makebox(0,0){$\circ$}}
\put(724,475){\makebox(0,0){$\circ$}}
\put(765,503){\makebox(0,0){$\circ$}}
\put(806,537){\makebox(0,0){$\circ$}}
\put(847,578){\makebox(0,0){$\circ$}}
\put(888,623){\makebox(0,0){$\circ$}}
\put(929,673){\makebox(0,0){$\circ$}}
\put(970,727){\makebox(0,0){$\circ$}}
\put(1011,785){\makebox(0,0){$\circ$}}
\put(777,700){\makebox(0,0){$\circ$}}
\put(707,659){\makebox(0,0)[r]{$Cmin$}}
\put(727.0,659.0){\rule[-0.200pt]{24.090pt}{0.400pt}}
\put(191,193){\usebox{\plotpoint}}
\multiput(191.00,191.92)(0.791,-0.497){49}{\rule{0.731pt}{0.120pt}}
\multiput(191.00,192.17)(39.483,-26.000){2}{\rule{0.365pt}{0.400pt}}
\multiput(232.00,165.92)(1.033,-0.496){37}{\rule{0.920pt}{0.119pt}}
\multiput(232.00,166.17)(39.090,-20.000){2}{\rule{0.460pt}{0.400pt}}
\multiput(273.00,145.92)(1.746,-0.492){21}{\rule{1.467pt}{0.119pt}}
\multiput(273.00,146.17)(37.956,-12.000){2}{\rule{0.733pt}{0.400pt}}
\multiput(314.00,133.94)(5.891,-0.468){5}{\rule{4.200pt}{0.113pt}}
\multiput(314.00,134.17)(32.283,-4.000){2}{\rule{2.100pt}{0.400pt}}
\multiput(355.00,131.60)(5.891,0.468){5}{\rule{4.200pt}{0.113pt}}
\multiput(355.00,130.17)(32.283,4.000){2}{\rule{2.100pt}{0.400pt}}
\multiput(396.00,135.58)(1.746,0.492){21}{\rule{1.467pt}{0.119pt}}
\multiput(396.00,134.17)(37.956,12.000){2}{\rule{0.733pt}{0.400pt}}
\multiput(437.00,147.58)(1.033,0.496){37}{\rule{0.920pt}{0.119pt}}
\multiput(437.00,146.17)(39.090,20.000){2}{\rule{0.460pt}{0.400pt}}
\multiput(478.00,167.58)(0.791,0.497){49}{\rule{0.731pt}{0.120pt}}
\multiput(478.00,166.17)(39.483,26.000){2}{\rule{0.365pt}{0.400pt}}
\multiput(519.00,193.58)(0.621,0.497){63}{\rule{0.597pt}{0.120pt}}
\multiput(519.00,192.17)(39.761,33.000){2}{\rule{0.298pt}{0.400pt}}
\multiput(560.00,226.58)(0.553,0.498){71}{\rule{0.543pt}{0.120pt}}
\multiput(560.00,225.17)(39.872,37.000){2}{\rule{0.272pt}{0.400pt}}
\multiput(601.58,263.00)(0.498,0.511){79}{\rule{0.120pt}{0.510pt}}
\multiput(600.17,263.00)(41.000,40.942){2}{\rule{0.400pt}{0.255pt}}
\multiput(642.58,305.00)(0.498,0.548){79}{\rule{0.120pt}{0.539pt}}
\multiput(641.17,305.00)(41.000,43.881){2}{\rule{0.400pt}{0.270pt}}
\multiput(683.58,350.00)(0.498,0.585){79}{\rule{0.120pt}{0.568pt}}
\multiput(682.17,350.00)(41.000,46.820){2}{\rule{0.400pt}{0.284pt}}
\multiput(724.58,398.00)(0.498,0.622){79}{\rule{0.120pt}{0.598pt}}
\multiput(723.17,398.00)(41.000,49.760){2}{\rule{0.400pt}{0.299pt}}
\multiput(765.58,449.00)(0.498,0.634){79}{\rule{0.120pt}{0.607pt}}
\multiput(764.17,449.00)(41.000,50.739){2}{\rule{0.400pt}{0.304pt}}
\multiput(806.58,501.00)(0.498,0.671){79}{\rule{0.120pt}{0.637pt}}
\multiput(805.17,501.00)(41.000,53.679){2}{\rule{0.400pt}{0.318pt}}
\multiput(847.58,556.00)(0.498,0.671){79}{\rule{0.120pt}{0.637pt}}
\multiput(846.17,556.00)(41.000,53.679){2}{\rule{0.400pt}{0.318pt}}
\multiput(888.58,611.00)(0.498,0.696){79}{\rule{0.120pt}{0.656pt}}
\multiput(887.17,611.00)(41.000,55.638){2}{\rule{0.400pt}{0.328pt}}
\multiput(929.58,668.00)(0.498,0.708){79}{\rule{0.120pt}{0.666pt}}
\multiput(928.17,668.00)(41.000,56.618){2}{\rule{0.400pt}{0.333pt}}
\multiput(970.58,726.00)(0.498,0.720){79}{\rule{0.120pt}{0.676pt}}
\multiput(969.17,726.00)(41.000,57.598){2}{\rule{0.400pt}{0.338pt}}
\put(191,193){\makebox(0,0){$\bullet$}}
\put(232,167){\makebox(0,0){$\bullet$}}
\put(273,147){\makebox(0,0){$\bullet$}}
\put(314,135){\makebox(0,0){$\bullet$}}
\put(355,131){\makebox(0,0){$\bullet$}}
\put(396,135){\makebox(0,0){$\bullet$}}
\put(437,147){\makebox(0,0){$\bullet$}}
\put(478,167){\makebox(0,0){$\bullet$}}
\put(519,193){\makebox(0,0){$\bullet$}}
\put(560,226){\makebox(0,0){$\bullet$}}
\put(601,263){\makebox(0,0){$\bullet$}}
\put(642,305){\makebox(0,0){$\bullet$}}
\put(683,350){\makebox(0,0){$\bullet$}}
\put(724,398){\makebox(0,0){$\bullet$}}
\put(765,449){\makebox(0,0){$\bullet$}}
\put(806,501){\makebox(0,0){$\bullet$}}
\put(847,556){\makebox(0,0){$\bullet$}}
\put(888,611){\makebox(0,0){$\bullet$}}
\put(929,668){\makebox(0,0){$\bullet$}}
\put(970,726){\makebox(0,0){$\bullet$}}
\put(1011,785){\makebox(0,0){$\bullet$}}
\put(777,659){\makebox(0,0){$\bullet$}}
\sbox{\plotpoint}{\rule[-0.400pt]{0.800pt}{0.800pt}}%
\sbox{\plotpoint}{\rule[-0.200pt]{0.400pt}{0.400pt}}%
\put(707,618){\makebox(0,0)[r]{$S vs C$}}
\sbox{\plotpoint}{\rule[-0.400pt]{0.800pt}{0.800pt}}%
\put(532,240){\makebox(0,0){$\ast$}}
\put(353,289){\makebox(0,0){$\ast$}}
\put(444,206){\makebox(0,0){$\ast$}}
\put(366,445){\makebox(0,0){$\ast$}}
\put(298,245){\makebox(0,0){$\ast$}}
\put(672,408){\makebox(0,0){$\ast$}}
\put(503,198){\makebox(0,0){$\ast$}}
\put(436,199){\makebox(0,0){$\ast$}}
\put(427,196){\makebox(0,0){$\ast$}}
\put(367,217){\makebox(0,0){$\ast$}}
\put(244,289){\makebox(0,0){$\ast$}}
\put(582,347){\makebox(0,0){$\ast$}}
\put(261,312){\makebox(0,0){$\ast$}}
\put(334,346){\makebox(0,0){$\ast$}}
\put(317,353){\makebox(0,0){$\ast$}}
\put(472,271){\makebox(0,0){$\ast$}}
\put(290,390){\makebox(0,0){$\ast$}}
\put(315,182){\makebox(0,0){$\ast$}}
\put(386,255){\makebox(0,0){$\ast$}}
\put(285,218){\makebox(0,0){$\ast$}}
\put(324,221){\makebox(0,0){$\ast$}}
\put(371,222){\makebox(0,0){$\ast$}}
\put(295,205){\makebox(0,0){$\ast$}}
\put(555,266){\makebox(0,0){$\ast$}}
\put(312,179){\makebox(0,0){$\ast$}}
\put(233,219){\makebox(0,0){$\ast$}}
\put(469,321){\makebox(0,0){$\ast$}}
\put(285,261){\makebox(0,0){$\ast$}}
\put(516,272){\makebox(0,0){$\ast$}}
\put(266,432){\makebox(0,0){$\ast$}}
\put(777,618){\makebox(0,0){$\ast$}}
\sbox{\plotpoint}{\rule[-0.200pt]{0.400pt}{0.400pt}}%
\put(191.0,131.0){\rule[-0.200pt]{0.400pt}{157.549pt}}
\put(191.0,131.0){\rule[-0.200pt]{197.538pt}{0.400pt}}
\put(1011.0,131.0){\rule[-0.200pt]{0.400pt}{157.549pt}}
\put(191.0,785.0){\rule[-0.200pt]{197.538pt}{0.400pt}}
\end{picture}

\caption{Análisis de concentración de todas las preguntas del FCI.}
\label{Concen}
\end{center}
\end {figure}
Del análisis de concentración contra puntaje mostrado en la Fig.\ref{Concen} se puede deducir que el $53,3\%$ de las preguntas se respondieron sin modelo mental alguno, es decir al azar, un $34,4\%$ con predominio de dos modelos erróneos y tan sólo un $13,3\%$ de las preguntas se respondieron con la mezcla de un  modelo correcto y uno  incorrecto. Así mismo, se puede ver que  el $96,7\%$ de las preguntas se respondió con un puntaje menor a $0,5$ lo que implica que mucho más de la mitad de la muestra respondió de manera incorrecta a más de la mitad de las  preguntas.\\ 

\begin {figure}
\begin{center}
\setlength{\unitlength}{0.240900pt}
\ifx\plotpoint\undefined\newsavebox{\plotpoint}\fi
\sbox{\plotpoint}{\rule[-0.200pt]{0.400pt}{0.400pt}}%
\begin{picture}(1062,826)(0,0)
\sbox{\plotpoint}{\rule[-0.200pt]{0.400pt}{0.400pt}}%
\put(211.0,131.0){\rule[-0.200pt]{192.720pt}{0.400pt}}
\put(211.0,131.0){\rule[-0.200pt]{4.818pt}{0.400pt}}
\put(191,131){\makebox(0,0)[r]{ 0}}
\put(991.0,131.0){\rule[-0.200pt]{4.818pt}{0.400pt}}
\put(211.0,295.0){\rule[-0.200pt]{192.720pt}{0.400pt}}
\put(211.0,295.0){\rule[-0.200pt]{4.818pt}{0.400pt}}
\put(191,295){\makebox(0,0)[r]{ 0.05}}
\put(991.0,295.0){\rule[-0.200pt]{4.818pt}{0.400pt}}
\put(211.0,458.0){\rule[-0.200pt]{192.720pt}{0.400pt}}
\put(211.0,458.0){\rule[-0.200pt]{4.818pt}{0.400pt}}
\put(191,458){\makebox(0,0)[r]{ 0.1}}
\put(991.0,458.0){\rule[-0.200pt]{4.818pt}{0.400pt}}
\put(211.0,622.0){\rule[-0.200pt]{192.720pt}{0.400pt}}
\put(211.0,622.0){\rule[-0.200pt]{4.818pt}{0.400pt}}
\put(191,622){\makebox(0,0)[r]{ 0.15}}
\put(991.0,622.0){\rule[-0.200pt]{4.818pt}{0.400pt}}
\put(211.0,785.0){\rule[-0.200pt]{192.720pt}{0.400pt}}
\put(211.0,785.0){\rule[-0.200pt]{4.818pt}{0.400pt}}
\put(191,785){\makebox(0,0)[r]{ 0.2}}
\put(991.0,785.0){\rule[-0.200pt]{4.818pt}{0.400pt}}
\put(211.0,131.0){\rule[-0.200pt]{0.400pt}{157.549pt}}
\put(211.0,131.0){\rule[-0.200pt]{0.400pt}{4.818pt}}
\put(211,90){\makebox(0,0){$0$}}
\put(211.0,765.0){\rule[-0.200pt]{0.400pt}{4.818pt}}
\put(344.0,131.0){\rule[-0.200pt]{0.400pt}{157.549pt}}
\put(344.0,131.0){\rule[-0.200pt]{0.400pt}{4.818pt}}
\put(344,90){\makebox(0,0){$5$}}
\put(344.0,765.0){\rule[-0.200pt]{0.400pt}{4.818pt}}
\put(478.0,131.0){\rule[-0.200pt]{0.400pt}{157.549pt}}
\put(478.0,131.0){\rule[-0.200pt]{0.400pt}{4.818pt}}
\put(478,90){\makebox(0,0){$10$}}
\put(478.0,765.0){\rule[-0.200pt]{0.400pt}{4.818pt}}
\put(611.0,131.0){\rule[-0.200pt]{0.400pt}{157.549pt}}
\put(611.0,131.0){\rule[-0.200pt]{0.400pt}{4.818pt}}
\put(611,90){\makebox(0,0){$15$}}
\put(611.0,765.0){\rule[-0.200pt]{0.400pt}{4.818pt}}
\put(744.0,131.0){\rule[-0.200pt]{0.400pt}{132.013pt}}
\put(744.0,720.0){\rule[-0.200pt]{0.400pt}{15.658pt}}
\put(744.0,131.0){\rule[-0.200pt]{0.400pt}{4.818pt}}
\put(744,90){\makebox(0,0){$20$}}
\put(744.0,765.0){\rule[-0.200pt]{0.400pt}{4.818pt}}
\put(878.0,131.0){\rule[-0.200pt]{0.400pt}{132.013pt}}
\put(878.0,720.0){\rule[-0.200pt]{0.400pt}{15.658pt}}
\put(878.0,131.0){\rule[-0.200pt]{0.400pt}{4.818pt}}
\put(878,90){\makebox(0,0){$25$}}
\put(878.0,765.0){\rule[-0.200pt]{0.400pt}{4.818pt}}
\put(1011.0,131.0){\rule[-0.200pt]{0.400pt}{157.549pt}}
\put(1011.0,131.0){\rule[-0.200pt]{0.400pt}{4.818pt}}
\put(1011,90){\makebox(0,0){$30$}}
\put(1011.0,765.0){\rule[-0.200pt]{0.400pt}{4.818pt}}
\put(211.0,131.0){\rule[-0.200pt]{0.400pt}{157.549pt}}
\put(211.0,131.0){\rule[-0.200pt]{192.720pt}{0.400pt}}
\put(1011.0,131.0){\rule[-0.200pt]{0.400pt}{157.549pt}}
\put(211.0,785.0){\rule[-0.200pt]{192.720pt}{0.400pt}}
\put(50,458){\makebox(0,0){$\rotatebox{90} { \parbox{4cm}{\begin{align*}  f_{R}  \end{align*}}}$}}
\put(611,29){\makebox(0,0){$x$}}
\put(844,700){\makebox(0,0)[r]{$f_{R}$}}
\put(864.0,700.0){\rule[-0.200pt]{24.090pt}{0.400pt}}
\put(211,138){\usebox{\plotpoint}}
\multiput(211.00,138.59)(1.543,0.489){15}{\rule{1.300pt}{0.118pt}}
\multiput(211.00,137.17)(24.302,9.000){2}{\rule{0.650pt}{0.400pt}}
\multiput(238.58,147.00)(0.497,2.176){49}{\rule{0.120pt}{1.823pt}}
\multiput(237.17,147.00)(26.000,108.216){2}{\rule{0.400pt}{0.912pt}}
\multiput(264.58,259.00)(0.497,1.080){51}{\rule{0.120pt}{0.959pt}}
\multiput(263.17,259.00)(27.000,56.009){2}{\rule{0.400pt}{0.480pt}}
\multiput(291.58,317.00)(0.497,2.019){51}{\rule{0.120pt}{1.700pt}}
\multiput(290.17,317.00)(27.000,104.472){2}{\rule{0.400pt}{0.850pt}}
\multiput(318.58,425.00)(0.497,2.741){49}{\rule{0.120pt}{2.269pt}}
\multiput(317.17,425.00)(26.000,136.290){2}{\rule{0.400pt}{1.135pt}}
\multiput(344.58,566.00)(0.497,1.212){51}{\rule{0.120pt}{1.063pt}}
\multiput(343.17,566.00)(27.000,62.794){2}{\rule{0.400pt}{0.531pt}}
\multiput(371.58,625.97)(0.497,-1.399){51}{\rule{0.120pt}{1.211pt}}
\multiput(370.17,628.49)(27.000,-72.486){2}{\rule{0.400pt}{0.606pt}}
\multiput(398.58,553.09)(0.497,-0.752){49}{\rule{0.120pt}{0.700pt}}
\multiput(397.17,554.55)(26.000,-37.547){2}{\rule{0.400pt}{0.350pt}}
\multiput(424.58,509.76)(0.497,-2.075){51}{\rule{0.120pt}{1.744pt}}
\multiput(423.17,513.38)(27.000,-107.379){2}{\rule{0.400pt}{0.872pt}}
\multiput(451.58,401.34)(0.497,-1.287){51}{\rule{0.120pt}{1.122pt}}
\multiput(450.17,403.67)(27.000,-66.671){2}{\rule{0.400pt}{0.561pt}}
\multiput(478.58,330.96)(0.497,-1.708){49}{\rule{0.120pt}{1.454pt}}
\multiput(477.17,333.98)(26.000,-84.982){2}{\rule{0.400pt}{0.727pt}}
\multiput(504.00,247.92)(0.518,-0.497){49}{\rule{0.515pt}{0.120pt}}
\multiput(504.00,248.17)(25.930,-26.000){2}{\rule{0.258pt}{0.400pt}}
\multiput(531.00,223.61)(5.820,0.447){3}{\rule{3.700pt}{0.108pt}}
\multiput(531.00,222.17)(19.320,3.000){2}{\rule{1.850pt}{0.400pt}}
\multiput(558.58,223.67)(0.497,-0.576){49}{\rule{0.120pt}{0.562pt}}
\multiput(557.17,224.83)(26.000,-28.834){2}{\rule{0.400pt}{0.281pt}}
\multiput(584.58,192.82)(0.497,-0.836){51}{\rule{0.120pt}{0.767pt}}
\multiput(583.17,194.41)(27.000,-43.409){2}{\rule{0.400pt}{0.383pt}}
\multiput(611.00,149.92)(1.381,-0.491){17}{\rule{1.180pt}{0.118pt}}
\multiput(611.00,150.17)(24.551,-10.000){2}{\rule{0.590pt}{0.400pt}}
\multiput(638.00,139.95)(5.597,-0.447){3}{\rule{3.567pt}{0.108pt}}
\multiput(638.00,140.17)(18.597,-3.000){2}{\rule{1.783pt}{0.400pt}}
\multiput(664.00,138.61)(5.820,0.447){3}{\rule{3.700pt}{0.108pt}}
\multiput(664.00,137.17)(19.320,3.000){2}{\rule{1.850pt}{0.400pt}}
\multiput(691.00,139.92)(1.381,-0.491){17}{\rule{1.180pt}{0.118pt}}
\multiput(691.00,140.17)(24.551,-10.000){2}{\rule{0.590pt}{0.400pt}}
\put(211,138){\makebox(0,0){$\bullet$}}
\put(238,147){\makebox(0,0){$\bullet$}}
\put(264,259){\makebox(0,0){$\bullet$}}
\put(291,317){\makebox(0,0){$\bullet$}}
\put(318,425){\makebox(0,0){$\bullet$}}
\put(344,566){\makebox(0,0){$\bullet$}}
\put(371,631){\makebox(0,0){$\bullet$}}
\put(398,556){\makebox(0,0){$\bullet$}}
\put(424,517){\makebox(0,0){$\bullet$}}
\put(451,406){\makebox(0,0){$\bullet$}}
\put(478,337){\makebox(0,0){$\bullet$}}
\put(504,249){\makebox(0,0){$\bullet$}}
\put(531,223){\makebox(0,0){$\bullet$}}
\put(558,226){\makebox(0,0){$\bullet$}}
\put(584,196){\makebox(0,0){$\bullet$}}
\put(611,151){\makebox(0,0){$\bullet$}}
\put(638,141){\makebox(0,0){$\bullet$}}
\put(664,138){\makebox(0,0){$\bullet$}}
\put(691,141){\makebox(0,0){$\bullet$}}
\put(718,131){\makebox(0,0){$\bullet$}}
\put(744,131){\makebox(0,0){$\bullet$}}
\put(771,131){\makebox(0,0){$\bullet$}}
\put(798,131){\makebox(0,0){$\bullet$}}
\put(824,131){\makebox(0,0){$\bullet$}}
\put(851,131){\makebox(0,0){$\bullet$}}
\put(878,131){\makebox(0,0){$\bullet$}}
\put(904,131){\makebox(0,0){$\bullet$}}
\put(931,131){\makebox(0,0){$\bullet$}}
\put(958,131){\makebox(0,0){$\bullet$}}
\put(984,131){\makebox(0,0){$\bullet$}}
\put(1011,131){\makebox(0,0){$\bullet$}}
\put(914,700){\makebox(0,0){$\bullet$}}
\put(718.0,131.0){\rule[-0.200pt]{70.584pt}{0.400pt}}
\put(211.0,131.0){\rule[-0.200pt]{0.400pt}{157.549pt}}
\put(211.0,131.0){\rule[-0.200pt]{192.720pt}{0.400pt}}
\put(1011.0,131.0){\rule[-0.200pt]{0.400pt}{157.549pt}}
\put(211.0,785.0){\rule[-0.200pt]{192.720pt}{0.400pt}}
\end{picture}

\caption{Frecuencia relativa del número de preguntas correctas.}
\label{FR}
\end{center}
\end {figure}
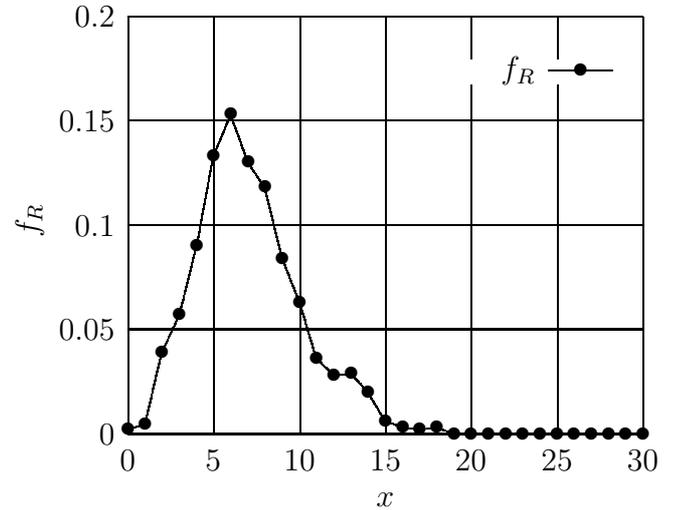

De otra parte la frecuencia relativa $f_R$ del número de preguntas correctamente respondidas se muestra en la Fig.\ref{FR}. Aquí se observa que el número de preguntas correctas que la muestra responde con mayor frecuencia es $6$ y que más de $15$ preguntas correctas no son respondidas. También se observa que $f_R$ es parecida una distribución de probabilidad binomial con media $6$ y con probabilidad de éxito de $0,2$, lo que permite entender al menos en parte el comportamiento aleatorio mostrado en el análisis de concentración, ver Fig.\ref{Concen}.\\ 
\section{Modelo aleatorio de dos niveles}
De acuerdo con la sección anterior los estudiantes  en estudio no manifiestan de manera significativa modelos mentales correctos sobre los conceptos físicos asociada al mundo Newtoniano, sino que su enfoque de respuestas es aleatorio con pocos sesgos hacia las ideas comunes y mucho menos  hacia las ideas físicamente correctas.\\

El modelo aleatorio de dos niveles consiste en lo siguiente: dados $N_e$ estudiantes una proporción $N_{inf}$ responden el cuestionario con probabilidad de acierto $p_{inf}$ en cada pregunta y otra  proporción  $N_{sup}$ responden el cuestionario con probabilidad $p_{sup}$ de acierto en cada pregunta, es decir el modelo considera que hay sólo dos tipos de estudiantes cada uno con niveles diferentes de responder correctamente.\\

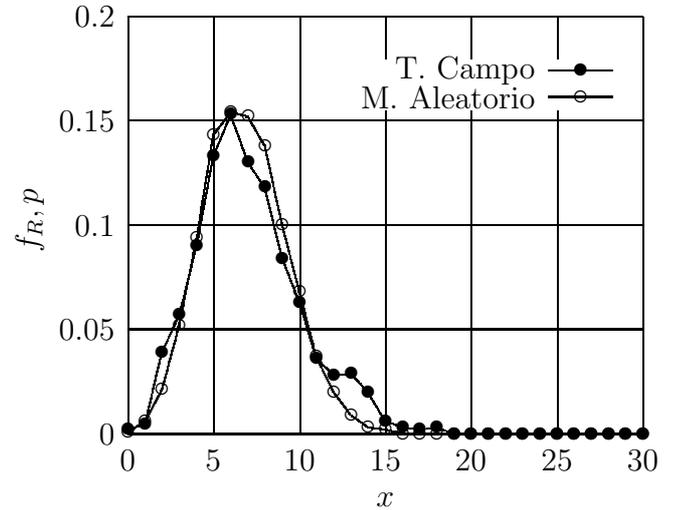
\begin {figure}
\begin{center}
\setlength{\unitlength}{0.240900pt}
\ifx\plotpoint\undefined\newsavebox{\plotpoint}\fi
\sbox{\plotpoint}{\rule[-0.200pt]{0.400pt}{0.400pt}}%
\begin{picture}(1062,826)(0,0)
\sbox{\plotpoint}{\rule[-0.200pt]{0.400pt}{0.400pt}}%
\put(211.0,131.0){\rule[-0.200pt]{192.720pt}{0.400pt}}
\put(211.0,131.0){\rule[-0.200pt]{4.818pt}{0.400pt}}
\put(191,131){\makebox(0,0)[r]{ 0}}
\put(991.0,131.0){\rule[-0.200pt]{4.818pt}{0.400pt}}
\put(211.0,295.0){\rule[-0.200pt]{192.720pt}{0.400pt}}
\put(211.0,295.0){\rule[-0.200pt]{4.818pt}{0.400pt}}
\put(191,295){\makebox(0,0)[r]{ 0.05}}
\put(991.0,295.0){\rule[-0.200pt]{4.818pt}{0.400pt}}
\put(211.0,458.0){\rule[-0.200pt]{192.720pt}{0.400pt}}
\put(211.0,458.0){\rule[-0.200pt]{4.818pt}{0.400pt}}
\put(191,458){\makebox(0,0)[r]{ 0.1}}
\put(991.0,458.0){\rule[-0.200pt]{4.818pt}{0.400pt}}
\put(211.0,622.0){\rule[-0.200pt]{192.720pt}{0.400pt}}
\put(211.0,622.0){\rule[-0.200pt]{4.818pt}{0.400pt}}
\put(191,622){\makebox(0,0)[r]{ 0.15}}
\put(991.0,622.0){\rule[-0.200pt]{4.818pt}{0.400pt}}
\put(211.0,785.0){\rule[-0.200pt]{192.720pt}{0.400pt}}
\put(211.0,785.0){\rule[-0.200pt]{4.818pt}{0.400pt}}
\put(191,785){\makebox(0,0)[r]{ 0.2}}
\put(991.0,785.0){\rule[-0.200pt]{4.818pt}{0.400pt}}
\put(211.0,131.0){\rule[-0.200pt]{0.400pt}{157.549pt}}
\put(211.0,131.0){\rule[-0.200pt]{0.400pt}{4.818pt}}
\put(211,90){\makebox(0,0){$0$}}
\put(211.0,765.0){\rule[-0.200pt]{0.400pt}{4.818pt}}
\put(344.0,131.0){\rule[-0.200pt]{0.400pt}{157.549pt}}
\put(344.0,131.0){\rule[-0.200pt]{0.400pt}{4.818pt}}
\put(344,90){\makebox(0,0){$5$}}
\put(344.0,765.0){\rule[-0.200pt]{0.400pt}{4.818pt}}
\put(478.0,131.0){\rule[-0.200pt]{0.400pt}{157.549pt}}
\put(478.0,131.0){\rule[-0.200pt]{0.400pt}{4.818pt}}
\put(478,90){\makebox(0,0){$10$}}
\put(478.0,765.0){\rule[-0.200pt]{0.400pt}{4.818pt}}
\put(611.0,131.0){\rule[-0.200pt]{0.400pt}{122.136pt}}
\put(611.0,720.0){\rule[-0.200pt]{0.400pt}{15.658pt}}
\put(611.0,131.0){\rule[-0.200pt]{0.400pt}{4.818pt}}
\put(611,90){\makebox(0,0){$15$}}
\put(611.0,765.0){\rule[-0.200pt]{0.400pt}{4.818pt}}
\put(744.0,131.0){\rule[-0.200pt]{0.400pt}{122.136pt}}
\put(744.0,720.0){\rule[-0.200pt]{0.400pt}{15.658pt}}
\put(744.0,131.0){\rule[-0.200pt]{0.400pt}{4.818pt}}
\put(744,90){\makebox(0,0){$20$}}
\put(744.0,765.0){\rule[-0.200pt]{0.400pt}{4.818pt}}
\put(878.0,131.0){\rule[-0.200pt]{0.400pt}{122.136pt}}
\put(878.0,720.0){\rule[-0.200pt]{0.400pt}{15.658pt}}
\put(878.0,131.0){\rule[-0.200pt]{0.400pt}{4.818pt}}
\put(878,90){\makebox(0,0){$25$}}
\put(878.0,765.0){\rule[-0.200pt]{0.400pt}{4.818pt}}
\put(1011.0,131.0){\rule[-0.200pt]{0.400pt}{157.549pt}}
\put(1011.0,131.0){\rule[-0.200pt]{0.400pt}{4.818pt}}
\put(1011,90){\makebox(0,0){$30$}}
\put(1011.0,765.0){\rule[-0.200pt]{0.400pt}{4.818pt}}
\put(211.0,131.0){\rule[-0.200pt]{0.400pt}{157.549pt}}
\put(211.0,131.0){\rule[-0.200pt]{192.720pt}{0.400pt}}
\put(1011.0,131.0){\rule[-0.200pt]{0.400pt}{157.549pt}}
\put(211.0,785.0){\rule[-0.200pt]{192.720pt}{0.400pt}}
\put(50,458){\makebox(0,0){$\rotatebox{90} { \parbox{4cm}{\begin{align*}  f_{R}, p  \end{align*}}}$}}
\put(611,29){\makebox(0,0){$x$}}
\put(844,700){\makebox(0,0)[r]{T. Campo}}
\put(864.0,700.0){\rule[-0.200pt]{24.090pt}{0.400pt}}
\put(211,138){\usebox{\plotpoint}}
\multiput(211.00,138.59)(1.543,0.489){15}{\rule{1.300pt}{0.118pt}}
\multiput(211.00,137.17)(24.302,9.000){2}{\rule{0.650pt}{0.400pt}}
\multiput(238.58,147.00)(0.497,2.176){49}{\rule{0.120pt}{1.823pt}}
\multiput(237.17,147.00)(26.000,108.216){2}{\rule{0.400pt}{0.912pt}}
\multiput(264.58,259.00)(0.497,1.080){51}{\rule{0.120pt}{0.959pt}}
\multiput(263.17,259.00)(27.000,56.009){2}{\rule{0.400pt}{0.480pt}}
\multiput(291.58,317.00)(0.497,2.019){51}{\rule{0.120pt}{1.700pt}}
\multiput(290.17,317.00)(27.000,104.472){2}{\rule{0.400pt}{0.850pt}}
\multiput(318.58,425.00)(0.497,2.741){49}{\rule{0.120pt}{2.269pt}}
\multiput(317.17,425.00)(26.000,136.290){2}{\rule{0.400pt}{1.135pt}}
\multiput(344.58,566.00)(0.497,1.212){51}{\rule{0.120pt}{1.063pt}}
\multiput(343.17,566.00)(27.000,62.794){2}{\rule{0.400pt}{0.531pt}}
\multiput(371.58,625.97)(0.497,-1.399){51}{\rule{0.120pt}{1.211pt}}
\multiput(370.17,628.49)(27.000,-72.486){2}{\rule{0.400pt}{0.606pt}}
\multiput(398.58,553.09)(0.497,-0.752){49}{\rule{0.120pt}{0.700pt}}
\multiput(397.17,554.55)(26.000,-37.547){2}{\rule{0.400pt}{0.350pt}}
\multiput(424.58,509.76)(0.497,-2.075){51}{\rule{0.120pt}{1.744pt}}
\multiput(423.17,513.38)(27.000,-107.379){2}{\rule{0.400pt}{0.872pt}}
\multiput(451.58,401.34)(0.497,-1.287){51}{\rule{0.120pt}{1.122pt}}
\multiput(450.17,403.67)(27.000,-66.671){2}{\rule{0.400pt}{0.561pt}}
\multiput(478.58,330.96)(0.497,-1.708){49}{\rule{0.120pt}{1.454pt}}
\multiput(477.17,333.98)(26.000,-84.982){2}{\rule{0.400pt}{0.727pt}}
\multiput(504.00,247.92)(0.518,-0.497){49}{\rule{0.515pt}{0.120pt}}
\multiput(504.00,248.17)(25.930,-26.000){2}{\rule{0.258pt}{0.400pt}}
\multiput(531.00,223.61)(5.820,0.447){3}{\rule{3.700pt}{0.108pt}}
\multiput(531.00,222.17)(19.320,3.000){2}{\rule{1.850pt}{0.400pt}}
\multiput(558.58,223.67)(0.497,-0.576){49}{\rule{0.120pt}{0.562pt}}
\multiput(557.17,224.83)(26.000,-28.834){2}{\rule{0.400pt}{0.281pt}}
\multiput(584.58,192.82)(0.497,-0.836){51}{\rule{0.120pt}{0.767pt}}
\multiput(583.17,194.41)(27.000,-43.409){2}{\rule{0.400pt}{0.383pt}}
\multiput(611.00,149.92)(1.381,-0.491){17}{\rule{1.180pt}{0.118pt}}
\multiput(611.00,150.17)(24.551,-10.000){2}{\rule{0.590pt}{0.400pt}}
\multiput(638.00,139.95)(5.597,-0.447){3}{\rule{3.567pt}{0.108pt}}
\multiput(638.00,140.17)(18.597,-3.000){2}{\rule{1.783pt}{0.400pt}}
\multiput(664.00,138.61)(5.820,0.447){3}{\rule{3.700pt}{0.108pt}}
\multiput(664.00,137.17)(19.320,3.000){2}{\rule{1.850pt}{0.400pt}}
\multiput(691.00,139.92)(1.381,-0.491){17}{\rule{1.180pt}{0.118pt}}
\multiput(691.00,140.17)(24.551,-10.000){2}{\rule{0.590pt}{0.400pt}}
\put(211,138){\makebox(0,0){$\bullet$}}
\put(238,147){\makebox(0,0){$\bullet$}}
\put(264,259){\makebox(0,0){$\bullet$}}
\put(291,317){\makebox(0,0){$\bullet$}}
\put(318,425){\makebox(0,0){$\bullet$}}
\put(344,566){\makebox(0,0){$\bullet$}}
\put(371,631){\makebox(0,0){$\bullet$}}
\put(398,556){\makebox(0,0){$\bullet$}}
\put(424,517){\makebox(0,0){$\bullet$}}
\put(451,406){\makebox(0,0){$\bullet$}}
\put(478,337){\makebox(0,0){$\bullet$}}
\put(504,249){\makebox(0,0){$\bullet$}}
\put(531,223){\makebox(0,0){$\bullet$}}
\put(558,226){\makebox(0,0){$\bullet$}}
\put(584,196){\makebox(0,0){$\bullet$}}
\put(611,151){\makebox(0,0){$\bullet$}}
\put(638,141){\makebox(0,0){$\bullet$}}
\put(664,138){\makebox(0,0){$\bullet$}}
\put(691,141){\makebox(0,0){$\bullet$}}
\put(718,131){\makebox(0,0){$\bullet$}}
\put(744,131){\makebox(0,0){$\bullet$}}
\put(771,131){\makebox(0,0){$\bullet$}}
\put(798,131){\makebox(0,0){$\bullet$}}
\put(824,131){\makebox(0,0){$\bullet$}}
\put(851,131){\makebox(0,0){$\bullet$}}
\put(878,131){\makebox(0,0){$\bullet$}}
\put(904,131){\makebox(0,0){$\bullet$}}
\put(931,131){\makebox(0,0){$\bullet$}}
\put(958,131){\makebox(0,0){$\bullet$}}
\put(984,131){\makebox(0,0){$\bullet$}}
\put(1011,131){\makebox(0,0){$\bullet$}}
\put(914,700){\makebox(0,0){$\bullet$}}
\put(718.0,131.0){\rule[-0.200pt]{70.584pt}{0.400pt}}
\put(844,659){\makebox(0,0)[r]{M. Aleatorio}}
\put(864.0,659.0){\rule[-0.200pt]{24.090pt}{0.400pt}}
\put(211,134){\usebox{\plotpoint}}
\multiput(211.00,134.58)(0.798,0.495){31}{\rule{0.735pt}{0.119pt}}
\multiput(211.00,133.17)(25.474,17.000){2}{\rule{0.368pt}{0.400pt}}
\multiput(238.58,151.00)(0.497,0.947){49}{\rule{0.120pt}{0.854pt}}
\multiput(237.17,151.00)(26.000,47.228){2}{\rule{0.400pt}{0.427pt}}
\multiput(264.58,200.00)(0.497,1.887){51}{\rule{0.120pt}{1.596pt}}
\multiput(263.17,200.00)(27.000,97.687){2}{\rule{0.400pt}{0.798pt}}
\multiput(291.58,301.00)(0.497,2.563){51}{\rule{0.120pt}{2.130pt}}
\multiput(290.17,301.00)(27.000,132.580){2}{\rule{0.400pt}{1.065pt}}
\multiput(318.58,438.00)(0.497,3.131){49}{\rule{0.120pt}{2.577pt}}
\multiput(317.17,438.00)(26.000,155.651){2}{\rule{0.400pt}{1.288pt}}
\multiput(344.58,599.00)(0.497,0.667){51}{\rule{0.120pt}{0.633pt}}
\multiput(343.17,599.00)(27.000,34.685){2}{\rule{0.400pt}{0.317pt}}
\multiput(371.00,633.93)(2.018,-0.485){11}{\rule{1.643pt}{0.117pt}}
\multiput(371.00,634.17)(23.590,-7.000){2}{\rule{0.821pt}{0.400pt}}
\multiput(398.58,624.65)(0.497,-0.888){49}{\rule{0.120pt}{0.808pt}}
\multiput(397.17,626.32)(26.000,-44.324){2}{\rule{0.400pt}{0.404pt}}
\multiput(424.58,573.96)(0.497,-2.319){51}{\rule{0.120pt}{1.937pt}}
\multiput(423.17,577.98)(27.000,-119.980){2}{\rule{0.400pt}{0.969pt}}
\multiput(451.58,451.13)(0.497,-1.963){51}{\rule{0.120pt}{1.656pt}}
\multiput(450.17,454.56)(27.000,-101.564){2}{\rule{0.400pt}{0.828pt}}
\multiput(478.58,346.13)(0.497,-1.961){49}{\rule{0.120pt}{1.654pt}}
\multiput(477.17,349.57)(26.000,-97.567){2}{\rule{0.400pt}{0.827pt}}
\multiput(504.58,248.14)(0.497,-1.043){51}{\rule{0.120pt}{0.930pt}}
\multiput(503.17,250.07)(27.000,-54.071){2}{\rule{0.400pt}{0.465pt}}
\multiput(531.58,193.37)(0.497,-0.667){51}{\rule{0.120pt}{0.633pt}}
\multiput(530.17,194.69)(27.000,-34.685){2}{\rule{0.400pt}{0.317pt}}
\multiput(558.00,158.92)(0.686,-0.495){35}{\rule{0.647pt}{0.119pt}}
\multiput(558.00,159.17)(24.656,-19.000){2}{\rule{0.324pt}{0.400pt}}
\multiput(584.00,139.95)(5.820,-0.447){3}{\rule{3.700pt}{0.108pt}}
\multiput(584.00,140.17)(19.320,-3.000){2}{\rule{1.850pt}{0.400pt}}
\multiput(611.00,136.93)(2.018,-0.485){11}{\rule{1.643pt}{0.117pt}}
\multiput(611.00,137.17)(23.590,-7.000){2}{\rule{0.821pt}{0.400pt}}
\put(211,134){\makebox(0,0){$\circ$}}
\put(238,151){\makebox(0,0){$\circ$}}
\put(264,200){\makebox(0,0){$\circ$}}
\put(291,301){\makebox(0,0){$\circ$}}
\put(318,438){\makebox(0,0){$\circ$}}
\put(344,599){\makebox(0,0){$\circ$}}
\put(371,635){\makebox(0,0){$\circ$}}
\put(398,628){\makebox(0,0){$\circ$}}
\put(424,582){\makebox(0,0){$\circ$}}
\put(451,458){\makebox(0,0){$\circ$}}
\put(478,353){\makebox(0,0){$\circ$}}
\put(504,252){\makebox(0,0){$\circ$}}
\put(531,196){\makebox(0,0){$\circ$}}
\put(558,160){\makebox(0,0){$\circ$}}
\put(584,141){\makebox(0,0){$\circ$}}
\put(611,138){\makebox(0,0){$\circ$}}
\put(638,131){\makebox(0,0){$\circ$}}
\put(664,131){\makebox(0,0){$\circ$}}
\put(691,131){\makebox(0,0){$\circ$}}
\put(718,131){\makebox(0,0){$\circ$}}
\put(744,131){\makebox(0,0){$\circ$}}
\put(771,131){\makebox(0,0){$\circ$}}
\put(798,131){\makebox(0,0){$\circ$}}
\put(824,131){\makebox(0,0){$\circ$}}
\put(851,131){\makebox(0,0){$\circ$}}
\put(878,131){\makebox(0,0){$\circ$}}
\put(904,131){\makebox(0,0){$\circ$}}
\put(931,131){\makebox(0,0){$\circ$}}
\put(958,131){\makebox(0,0){$\circ$}}
\put(984,131){\makebox(0,0){$\circ$}}
\put(1011,131){\makebox(0,0){$\circ$}}
\put(914,659){\makebox(0,0){$\circ$}}
\put(638.0,131.0){\rule[-0.200pt]{89.856pt}{0.400pt}}
\put(211.0,131.0){\rule[-0.200pt]{0.400pt}{157.549pt}}
\put(211.0,131.0){\rule[-0.200pt]{192.720pt}{0.400pt}}
\put(1011.0,131.0){\rule[-0.200pt]{0.400pt}{157.549pt}}
\put(211.0,785.0){\rule[-0.200pt]{192.720pt}{0.400pt}}
\end{picture}

\caption{Confrontación del modelo con los resultados del trabajo de campo.}
\label{Modelo}
\end{center}
\end {figure}

El modelo se materializó mediante la técnica de Montecarlo \cite{Landau}. Al ejecutarlo con los parámetros mostrados en la tabla \ref{tab1} se encontró que la frecuencia relativa $p$ simulada mediante Montecarlo tiene buen acuerdo con la obtenida en el trabajo de campo, en la Fig.\ref{Modelo}  se puede observar una máxima discrepancia entre el modelo y los datos de campo que corresponden aproximadamente al $2\%$ y se encuentra en el porcentaje de estudiantes que respondieron correctamente $7$ preguntas.\\

\begin{table}[!htp]
\caption{Parámetros}\label{tab1} 
\vspace{2mm}
\centering
\begin{tabular}{|c|c|c|c|c|}
  \hline
$N_{e}$ & $N_{inf}$ & $N_{sup}$ & $p_{inf}$& $p_{sup}$\\ \hline 
$10000$ & $0,53$  &  $0,47$   & $0,2$   & $0,25$      \\ \hline  
\end{tabular}
\end{table}
Cabe anotar que el planteamiento del modelo  tiene una formulación  equivalente a la siguiente: la muestra está compuesta por un $53\%$ de estudiantes que responden  sin modelo  alguno eligiendo  alguna de las cinco opciones con la misma probabilidad  y de un $47\%$ de estudiantes que con certeza descartan una opción o modelo  y los demás los responden al azar con igual probabilidad. Claro, este modelo describe muy aproximadamente el comportamiento real de la muestra y presenta algunas leves diferencias con los resultados de campo especialmente en $x=7$ y $x=13$, aún así los resultados son bastante confiables.

\section{Conclusiones}
Se presentó un modelo aleatorio de dos niveles que mostró un muy buen acuerdo con los resultados obtenidos al aplicar el FCI a $646$ estudiantes de ingeniería de Bogotá, de acuerdo con esto resulta claro que los estudiantes de la muestra respondieron el FCI fundamentalmente al azar. Esto  implica que para esta muestra el FCI no es un instrumento del todo confiable desde el punto de vista de la TCT.De otro lado, el modelo no permite  conocer la razón por la cual globalmente las respuestas tienen carácter aleatorio, esto podría darse ya sea por el completo desconocimiento de los fenómenos físicos o  por el desinterés al presentar la prueba, entre otras causas en principio desconocidas.\\

Las perspectivas de este modelo están enfocadas a la interpretación  de los resultados de una prueba tipo  FCI en términos de respuestas aleatorias de cada pregunta, así al controlar el número de niveles,  el porcentaje de la muestra en los diversos niveles y  la probabilidad de acierto asociada a cada nivel, se espera reproducir algunas frecuencias relativas extraídas directamente de los datos del trabajo de campo. Esto trae como consecuencia la hipótesis de que los estudiantes responden siempre al azar este tipo de cuestionarios pero en diversos niveles de probabilidad, además que una medida del desempeño es la cuantificación de la tendencia de la probabilidad a $1$.\\

\section*{Agradecimientos}
Los autores agradecen a la Facultada de Ingeniería y al Departamento de Ciencias Naturales de la Universidad  Central por el tiempo y los recursos asignados al proyecto de investigación: Medición de indicadores de aprendizaje de la física introductoria en estudiantes de ingeniería de universidades de Bogotá, la cual se  realizó durante el año $2012$. 

\renewcommand{\refname}{Referencias}

\end{document}